\documentclass[conference]{IEEEtran}
\IEEEoverridecommandlockouts
\usepackage{cite}
\usepackage{amsmath,amssymb,amsfonts}
\usepackage{algorithmic}
\usepackage{graphicx}
\usepackage{textcomp}
\usepackage{xcolor}

\usepackage{subfig}
\usepackage{hyperref}
\hypersetup{
colorlinks=true,
linkcolor=red,
anchorcolor=red,
citecolor=green,
urlcolor=black
}

\def\BibTeX{{\rm B\kern-.05em{\sc i\kern-.025em b}\kern-.08em
    T\kern-.1667em\lower.7ex\hbox{E}\kern-.125emX}}
\begin{document}

\title{Rate-Aware Learned Speech Compression
}

\author{
\IEEEauthorblockN{Jun Xu}
\IEEEauthorblockA{Institute of Image Communication\\
and Network Engineering,\\
Shanghai Jiao Tong University\\
xujunzz@sjtu.edu.cn}
\and
\IEEEauthorblockN{Zhengxue Cheng}
\IEEEauthorblockA{Institute of Image Communication\\
and Network Engineering,\\
Shanghai Jiao Tong University\\
zxcheng@sjtu.edu.cn}
\and
\IEEEauthorblockN{Guangchuan Chi}
\IEEEauthorblockA{Institute of Image Communication\\
and Network Engineering,\\
Shanghai Jiao Tong University\\
chimou0@sjtu.edu.cn}
\and
\IEEEauthorblockN{Yuhan Liu}
\IEEEauthorblockA{Institute of Image Communication\\
and Network Engineering,\\
Shanghai Jiao Tong University\\
liu1025221459@sjtu.edu.cn}
\and
\IEEEauthorblockN{Yuelin Hu}
\IEEEauthorblockA{Institute of Image Communication\\
and Network Engineering,\\
Shanghai Jiao Tong University\\
huyuelin51717221@sjtu.edu.cn}
\and
\IEEEauthorblockN{Li Song}
\IEEEauthorblockA{Institute of Image Communication\\
and Network Engineering,\\
Shanghai Jiao Tong University\\
song\_li@sjtu.edu.cn}
}

\maketitle

\begin{abstract}
The rapid rise of real-time communication and large language models has significantly increased the importance of speech compression.
Deep learning-based neural speech codecs have outperformed traditional signal-level speech codecs in terms of rate-distortion (RD) performance.
Typically, these neural codecs employ an encoder–quantizer–decoder architecture, where audio is first converted into latent code feature representations and then into discrete tokens.
However, this architecture exhibits insufficient RD performance due to two main drawbacks:
(1) the inadequate performance of the quantizer, challenging training processes, and issues such as codebook collapse;
(2) the limited representational capacity of the encoder and decoder, making it difficult to meet feature representation requirements across various bitrates.
In this paper, we propose a rate-aware learned speech compression scheme that replaces the quantizer with an advanced channel-wise entropy model to improve RD performance, simplify training, and avoid codebook collapse.
We employ multi-scale convolution and linear attention mixture blocks to enhance the representational capacity and flexibility of the encoder and decoder.
Experimental results demonstrate that the proposed method achieves state-of-the-art RD performance, obtaining 53.51\% BD-Rate bitrate saving in average, and achieves 0.26 BD-VisQol and 0.44 BD-PESQ gains.
\end{abstract}


\section{Introduction}

As a core component of real-time communication, speech data streams occupy a significant portion of internet traffic, highlighting the increasing importance of speech compression.
Traditional speech codecs like OPUS\cite{opus} and EVS\cite{evs} achieve efficient compression through meticulously engineered pipelines that integrate psycho-acoustic models and speech synthesis algorithms.
While these signal processing-based codecs provide high quality at relatively sufficient bitrates, their performance either degrades significantly or is unsupported at lower bitrates.
In recent years, deep learning-based neural speech codecs\cite{soundstream, encodec, funcodec, nips23} have demonstrated performance surpassing that of traditional codecs.
They exhibit considerable advantages at lower bitrates, which suggests their potential for narrow-band communication applications.

\begin{figure}[tbp]
    \centering
    \includegraphics[trim=2mm 3mm 2mm 0mm, clip, width=0.48 \textwidth]{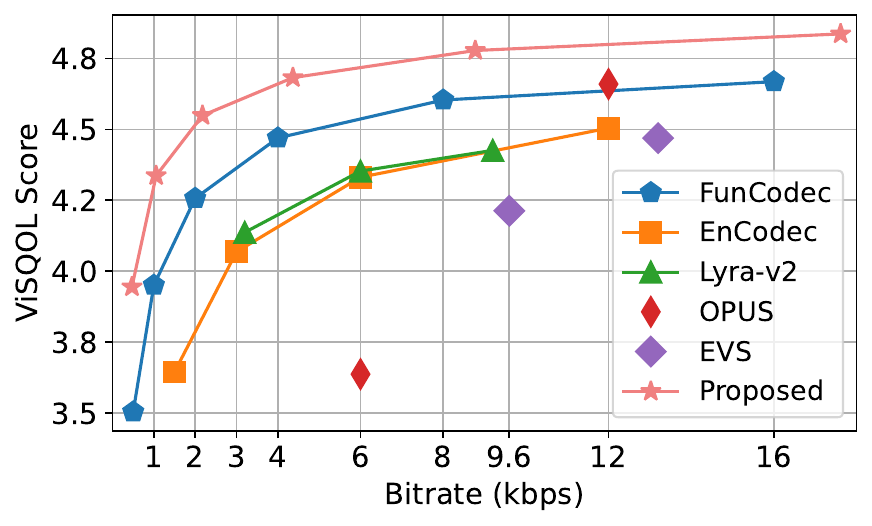}
    \caption{Comparison in terms of ViSQOL. The proposed scheme achieves 0.26 / -56.94\% in BD-ViSQOL / BD-RATE.}
    \label{teaser}
\vspace{-5mm}
\end{figure}

Neural speech codecs typically consist of an encoder, a quantizer and a decoder.
After pre-processing, the speech signal is input into the encoder, which performs feature extraction to obtain a continuous and compact representation in a high-dimensional latent space.
The quantizer then transforms this vector into discrete tokens in a lower-dimensional space.
The dequantizer and the decoder perform inverse operations of the quantizer and encoder, followed by post-processing to reconstruct the speech signal.

\begin{figure*}[htbp]
    \centering
    \includegraphics[trim=3mm 5.8mm 7mm 0.8mm, clip, width=1 \textwidth]{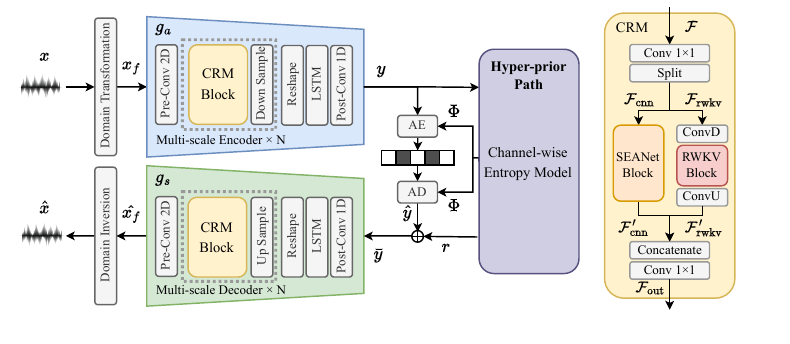}
    \caption{Architecture of the proposed Rate-Aware Learned Speech Compression, including the encoder, entropy model, and decoder. The encoder-decoder backbone utilizes CRM blocks which combine signal-level and non-local semantic information.}
    \label{architecture}
\vspace{-5mm}
\end{figure*}

In terms of encoder and decoder architecture, SoundStream\cite{soundstream} and EnCodec\cite{encodec} use SEANet\cite{seanet, seanet2} as the backbone, which is a convolution-based network.
EnCodec adds LSTM\cite{lstm} or Transformer\cite{transformer} layers at the end of the encoder for sequence modeling.
FunCodec\cite{funcodec} builds upon EnCodec by adding frequency-domain pre- and post-processing and extending SEANet into time-frequency domain.
Nevertheless, the encoder and decoder architectures of neural speech codecs still suffer from insufficient representational capacity and flexibility, which limits their rate-distortion (RD) performance.

Regarding the quantizer, neural speech codecs mostly use Residual Vector Quantization (RVQ)\cite{rvq, soundstream} to quantize the latent representations into discrete tokens.
RVQ achieves high efficient encoding by cascading multiple quantization layers.
This method allows the model to operate at different bitrates without the need to train separate models for each bitrate.
However, RVQ still faces issues like codebook collapse, leading to inefficient bit utilization.
\cite{nips23} alleviated this problem by using factorized codes and L2-normalized codes.
Language-Codec\cite{language-codec} employs Masked Channel RVQ to distribute more information across multiple quantization layers. 
Nevertheless, due to the difficulties in training optimization and codebook collapse problems, RVQ remains the main bottleneck limiting the performance of neural speech codecs.

To address the aforementioned issues and inspired by advancements in the field of learned image compression\cite{balle, cheng20, channel-wise, elic, tcm}, we propose using a rate-aware entropy model for the quantization module.
This approach not only leverages the well-established hyper-prior path to assist in compressing the latent representations, but also employs an channel-wise context model to enhance the accuracy of predicting the latent distribution.
Furthermore, in the encoder–decoder architecture, we introduce multi-scale convolution and attention\cite{rwkv, rwkv6} mixture (CRM) blocks as the backbone.
This design combines the local information captured by convolutions with the non-local correlations extracted by the attention mechanism at various scales, effectively improving the representational capacity and flexibility of encoder and decoder.
As shown in Fig. \ref{teaser}, our method achieves significant RD performance gains compared to schemes based on RVQ.

The contributions of this paper is summarized as follows:
\begin{itemize}
    \item We employ an advanced channel-wise entropy model and perform end-to-end training using a RD loss, which circumvents the training challenges and codebook collapse.
    \item We utilize CRM blocks to model both local and non-local features of the speech signal. This enhances the capacity and flexibility of the encoder and decoder.
    \item Experimental results show the proposed method achieves gains in two evaluation metrics, ViSQoL and PESQ.
\end{itemize}

\section{Proposed Method}

\begin{figure*}[htbp]
    \centering
    \includegraphics[trim=8mm 8mm 9mm 0, clip, width=1 \textwidth]{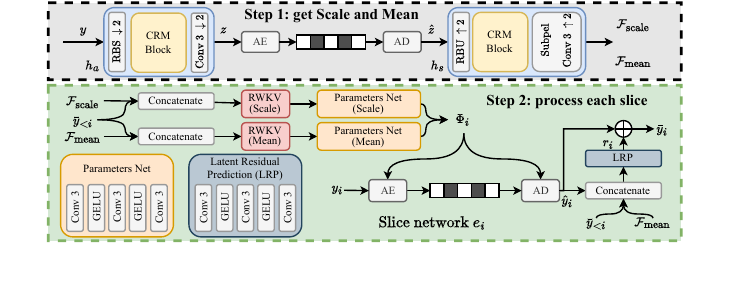}
    \caption{Architecture/steps of the entropy model. The first step is obtaining latent parameters from $y$. The second step processes each slice sequentially, using previously decoded slices to predict subsequent slices for more accurate reconstruction.}
    \label{entropy}
\vspace{-5mm}
\end{figure*}

\subsection{Overview of Learned Speech Compression}

As shown in Fig. \ref{architecture}, the proposed rate-aware learned speech compression model mainly consists of three components:
(1) An encoder network $g_a$ that takes the time-frequency representation $x_f$ of the preprocessed audio as input and outputs the latent representation $y$;
(2) A channel-wise entropy model that inputs the latent representation $y$ or its reconstruction and outputs the probability distribution $\Phi$ of $y$ and the residual $r$;
(3) A decoder network $g_s$ that takes the reconstruction $\bar{y}$ as input and outputs the reconstructed time-frequency representation $\hat{x}_f$ of the audio.

According to \cite{funcodec}, transforming the input speech signal into the frequency domain facilitates feature extraction and reconstruction.
Therefore, we use the Short-Time Fourier Transform (STFT) for domain transformation during pre-processing and post-processing.
The entire workflow can be formulated by:
\begin{equation}
    \begin{aligned}
        x_f &= \text{STFT}(x), \quad y=g_a(x_f ; \phi) \\
        \hat{y} &= Q(y-\mu)+\mu, \quad \bar{y}=\hat{y} \\
        \hat{x_f} &=g_s(\bar{y} ; \theta), \quad \hat{x}=\text{iSTFT}(\hat{x_f})
    \end{aligned}
    \label{eq1}
\end{equation}
where $x$ and $\hat{x}$ represent the raw audio and reconstructed audio, respectively, while $x_f$ and $\hat{x_f}$ are the time-frequency representation.
The encoder $g_a$ with parameters $\phi$ transforms the input $x_f$ into a latent representation $y$.
The quantization operation $Q$ quantizes $y$ to $\hat{y}$, where $\mu$ is the estimated mean.
The decoder $g_s$ with parameters $\theta$ reconstructs $\hat{x_f}$.
According to \cite{elic}, we round and encode $y - \mu$ to the bitstream instead of $y$ and restore the coding-symbol $\hat{y}$ as $Q(y - \mu) + \mu$, which facilitates entropy models. 
Additionally, the residual $r$ is utilized to reduce the quantization error ($y - \hat{y}$).
Consequently, the refined $\bar{y}$ is fed into the decoder $g_s$, rather than using $\hat{y}$.

\subsection{CNN-RWKV Mixture Block}

Current audio compression models are predominantly based on CNN and have achieved commendable RD performance.
Additionally, attention have proven advantageous across various tasks and domains because, unlike the inductive biases of CNN that focus on extracting local information, attention can capture non-local information.
In the context of feature extraction in the time-frequency domain of speech signals, both local and non-local information are crucial, emphasizing signal-level and semantic-level correlations, respectively.

To combine the strengths of CNN and attention, we propose the CNN-RWKV Mixture Block.
To support streaming input and maintain linear computational complexity, we select RWKV\cite{rwkv, rwkv6} as the attention module.
As shown on the right side of Fig. \ref{architecture}, the time-frequency features first pass through a 1×1 convolution and are then evenly split along the channel dimension into two parts.
These parts are separately fed into a fully convolutional SEANet block and an RWKV block.
Before the RWKV block's input and after its output, we downsample and upsample the input signals with convolution to balance computational cost and performance.
Subsequently, the two feature sets are concatenated, and a 1×1 convolution is used to fuse the features.

To capture local and non-local features of speech signals at different scales, we employ CRM blocks in conjunction with downsampling, as illustrated on the left side of Fig. \ref{architecture}.
As the degree of downsampling increases, high-level semantic information becomes more prominent, which we use more cascaded attention blocks to capture.
Moreover, due to the higher compactness of the samples, incorporating more cascaded blocks does not lead to excessive computational resource consumption.

\subsection{Channel-wise Entropy Model}

We adopt a channel-wise entropy model as Fig. \ref{entropy} shows, dividing $y$ into $s$ slices ${y_0, y_1, \cdots, y_{s-1}}$.
The encoding process for each slice is as follows:
\begin{equation}
    \begin{aligned}
        & z=h_a(y ; \phi_h), \quad \hat{z}=Q(z) \\
        & \mathcal{F}_{\text{mean}}, \mathcal{F}_{\text{scale}}=h_s(\hat{z} ; {\theta_h}) \\
        & r_i, \Phi_i=e_i(\mathcal{F}_{\text{mean}}, \mathcal{F}_{\text{scale}}, \bar{y}{<i}, y_i), 0 \leq i < s \\
        & \bar{y}_i=r_i+\hat{y}_i
    \end{aligned}
    \label{eq:entropy}
\end{equation}
where $h_a$ refers to the hyper-prior encoder, characterized by the parameters $\phi_h$.
This encoder provides the side information $z$, which is responsible for capturing correlations among the elements of $y$.
A factorized density model $\psi$ is employed to encode the quantized values $\hat{z}$ as
$
    p_{\hat{z} \mid \psi}(\hat{z} \mid \psi) = \prod_j \left(p_{z_j \mid \psi}(\psi) * \mathcal{U}\left(-\frac{1}{2}, \frac{1}{2}\right)\right)(\hat{z}_j)
$
where $j$ represents the index.
The quantized $\hat{z}$ is subsequently input into the hyper-prior decoder $h_s$, which is parameterized by $\theta_h$, to generate two latent features, $\mathcal{F}_{\text{mean}}$ and $\mathcal{F}_{\text{scale}}$.
In $h_a$ and $h_s$, we also  incorporated CRM blocks to achieve a more compact hyper-prior $z$ and more accurate estimation of latent parameters.
The above explains the meaning of the first two equations in Equation \ref{eq:entropy}, which also correspond to \textbf{Step 1} in Fig. \ref{entropy}.

\begin{figure*}[!t]
\centering
\subfloat[0.44 / -50.05\% in BD-PESQ / BD-RATE]{\includegraphics[trim=0mm 0mm 0mm 0mm, clip, width=0.49\textwidth]{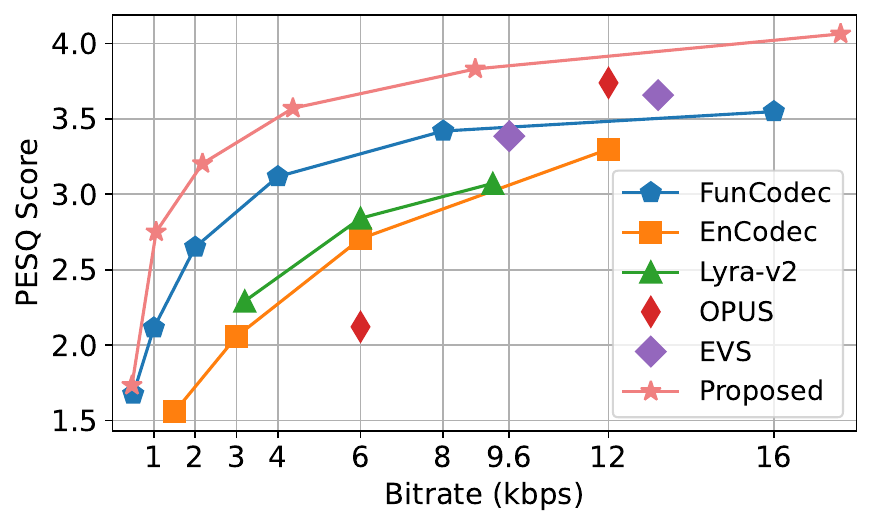}%
\label{fig_pesq}}
\subfloat[Rate-ViSQOL Curve in ablation study]{\includegraphics[trim=0mm 0mm 0mm 0mm, clip, width=0.49\textwidth]{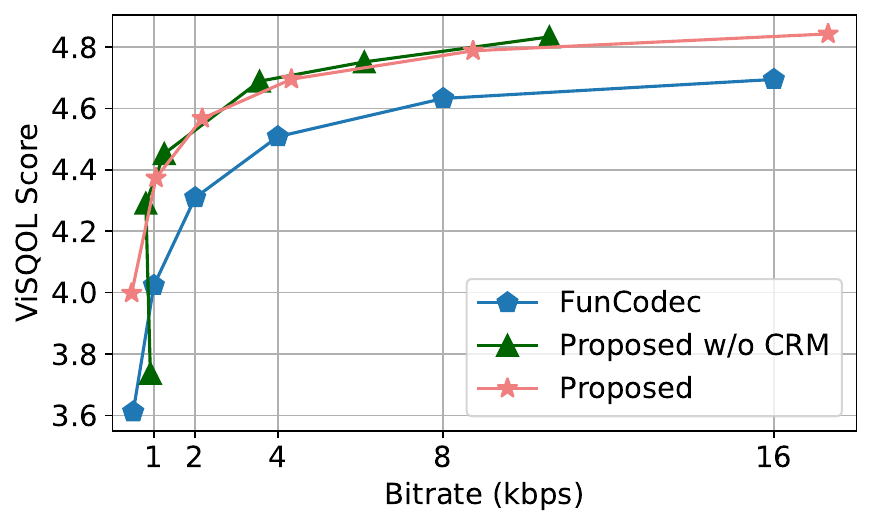}%
\label{fig_ablation}}
\caption{Subplots (a) shows the comparison between the proposed scheme and the baseline schemes in terms of the evaluation metrics ViSQOL. Subplot (b) presents the ViSQOL comparison in the ablation study.}
\label{fig_rd}
\vspace{-5mm}
\end{figure*}

Next, we explain the meaning of the last two equations, which correspond to \textbf{Step 2}.
The two latent features are then passed into each respective slice network $e_i$.
Following this, each slice $y_i$ is processed sequentially, resulting in $\bar{y}_i$.
During this sequence, the already encoded slices $\bar{y}_{<i} = \left\{\bar{y}_0, \bar{y}_1, \ldots, \bar{y}_{i-2}, \bar{y}_{i-1}\right\}$ along with the current slice $y_i$ are input to $e_i$, yielding the estimated distribution parameters $\Phi_i = \left(\mu_i, \sigma_i\right)$, which are necessary for generating bitstreams.
As a result, we assume that $p_{\hat{y} \mid \hat{z}}(\hat{y} \mid \hat{z}) \sim \mathcal{N}\left(\mu, \sigma^2\right)$.

Specifically, in $e_i$, $\mathcal{F}_{\text{scale}}$ and $\mathcal{F}_{\text{mean}}$ are concatenated with $\bar{y}_{<i}$, and then passed through an RWKV attention block and a convolutional Parameters Net to get $\Phi_i$.
Next, the decoded $\hat{y}_i$, the previous $\bar{y}_{<i}$, and $\mathcal{F}_{\text{mean}}$ are concatenated and fed into a Latent Residual Prediction network to estimate the residual $r_i$.
The final $\bar{y}_i$ is then obtained by adding the estimated residual $r_i$ to the decoded $\hat{y}_i$.

To train the learned speech compression model, we employ a Lagrangian multiplier-based rate-distortion optimization.
The loss function is defined as:
\begin{equation}
    \begin{aligned}
        \mathcal{L} &= \mathcal{R}(\hat{y}) + \mathcal{R}(\hat{z}) + \lambda \cdot \mathcal{D}(x, \hat{x}) \\
        &= \mathbb{E}[-\log_2(p_{\hat{y}|\hat{z}}(\hat{y}|\hat{z}))] + \mathbb{E}[-\log_2(p_{\hat{z}|\psi}(\hat{z}|\psi))] \\
        &\quad + \lambda \cdot \mathcal{D}(x, \hat{x})
    \end{aligned}
    \label{loss}
\end{equation}
where $\lambda$ controls the rate-distortion trade-off, $\mathcal{D}(x, \hat{x})$ is the distortion loss, and $\mathcal{R}(\hat{y})$ and $\mathcal{R}(\hat{z})$ represent the bit rates of the latents $\hat{y}$ and $\hat{z}$, respectively.

\section{Experiments}

\subsection{Experimental Setup}

\subsubsection{Dataset}
For our experiments, we utilized the LibriTTS corpus \cite{libritts} to train and evaluate the models.
We use the test-clean and test-other subset to evaluate the model, which contains high quality recordings, and includes more challenging audio conditions, respectively.

\subsubsection{Training Details}
The model is optimized using the rate-distortion function as the loss function, as shown in Equation \ref{loss}, where the distortion term includes both a time-domain reconstruction loss and a frequency-domain multi-scale Mel-spectrogram loss.
These two losses are defined as follows:
\begin{equation}
    \small
    \begin{aligned}
        \mathcal{L}_t(x, \hat{x}) &= \Vert x-\hat{x} \Vert_1\\
        \mathcal{L}_f(x, \hat{x}) &= \frac{1}{\| \alpha \|} \sum\limits_{i \in \alpha}(\Vert \mathcal{S}_i(x)-\mathcal{S}_i(\hat{x}) \Vert_1 + \Vert \mathcal{S}_i(x)-\mathcal{S}_i(\hat{x}) \Vert_2\\ &+ \Vert \mathcal{M}_i(x)-\mathcal{M}_i(\hat{x}) \Vert_1 + \Vert \mathcal{M}_i(x)-\mathcal{M}_i(\hat{x}) \Vert_2)\\
        \mathcal{D}(x, \hat{x}) &= \mathcal{L}_t(x, \hat{x}) + \mathcal{L}_f(x, \hat{x})
    \end{aligned}
    \label{distortion}
\end{equation}
where, $\mathcal{S}_i$ and $\mathcal{M}_i$ represent the log-compressed power and Mel spectra with a window size of $2^i$ and a shift length of $2^i/4$.
$\alpha$ is set to $[5, 6, ... ,11]$.
$\lambda$ belongs to $[0.25, 0.8, 2, 5.5, 9, 18]$ to get models with different bitrates.

\subsubsection{Baseline}
To evaluate the performance of our proposed model, we use two traditional speech codecs, OPUS\cite{opus} and EVS\cite{evs}, and three neural speech codecs, lyra-v2\cite{soundstream}, EnCodec\cite{encodec} and FunCodec\cite{funcodec}, as baselines.

\subsubsection{Evaluation Metrics}
We use ViSQOL\cite{visqol} and PESQ\cite{pesq} as metrics for speech quality.

\subsection{Experimental Results}

Fig. \ref{teaser} and \ref{fig_pesq} respectively illustrate the RD curves of the proposed speech compression scheme and the baseline scheme under two quality evaluation metrics, ViSQOL and PESQ, which demonstrating the superiority of the proposed scheme.
Compared to the baseline, the proposed scheme achieves improvements of 0.26 and 0.44 in BD-ViSQOL and BD-PESQ, while get -56.94\% and -50.05\% in BD-RATE, respectively.

\subsection{Ablation Study}

we proposes two major improvements, and Fig. \ref{fig_ablation} presents the results of ablation experiments on these improvements in the test-clean subset of LibriTTS.
The figure shows three RD curves corresponding to the baseline scheme, the proposed scheme, and the Entropy scheme (w/o CRM).
First, by comparing the baseline and Entropy scheme, it can be observed that the Entropy scheme achieves superior rate-distortion performance compared to the RVQ used in the baseline, outperforming it across the entire bitrate range.
The proposed Entropy module effectively addresses issues like codebook collapse inherent in RVQ.
Next, comparing the proposed complete scheme with the Entropy scheme (w/o CRM), it is evident that the CRM backbone provides flexibility at low bitrates for the overall neural coding architecture.

\section{Conclusion}

This paper proposes Rate-Aware Learned Speech Compression, an end-to-end audio compression scheme.
In the encoder-decoder architecture, multi-scale CNN and RWKV mixture blocks are used to enhance the backbone's representational capacity and flexibility.
A channel-wise entropy model replaces RVQ, achieving significant improvements in RD performance, allowing end-to-end training and avoiding codebook collapse issues.
Experimental results demonstrate the superiority of the proposed scheme, which has 53.51\% BD-Rate bitrate saving in average, 0.26 BD-VisQol and 0.44 BD-PESQ gains.

\newpage

\bibliographystyle{IEEEtran}
\bibliography{reference.bib}

\end{document}